\documentclass[%
 reprint,
 amsmath,amssymb,
 aps,
 prl,
prstab,
prstper,
floatfix,
]{revtex4-1}

\usepackage{graphicx}
\usepackage{dcolumn}
\usepackage{bm}
\usepackage{hyperref}
\usepackage[mathlines]{lineno}
\usepackage{lipsum}
\usepackage{amssymb}

\begin{document}



\title{Temperature-dependent stability of polytypes and stacking faults in SiC:\\
reconciling theory and experiments}

\author{Emilio Scalise}
\email{emilio.scalise@unimib.it}

\author{Anna Marzegalli}%
 
 \author{Francesco Montalenti}%
 
 \author{Leo Miglio}%
 
\affiliation{%
Department of Materials Science, University of Milano-Bicocca, Via Roberto Cozzi 55, 20125 Milan, IT
}%




\date{\today} 
\begin{abstract}
The relative stability of SiC polytypes, changing with temperature, has been considered a paradox for about thirty years, due to discrepancies between theory and experiments. Based on \textit{ab-initio} calculations including van der Waals corrections, a  temperature-dependent polytypic diagram consistent with the experimental observations is obtained. Results are easily interpreted based on the influence of the hexagonality on both cohesive energy and entropy. 
Temperature-dependent stability of stacking faults is also analyzed and found to be in agreement with experimental evidences. Our results suggest that lower temperatures during SiC crystal deposition are advantageous in order to reduce ubiquitous stacking faults in SiC-based power devices. 

\end{abstract}

\pacs{Valid PACS appear here}
\maketitle



\begin{figure}[b]
  \centering
    \includegraphics[width=0.48\textwidth]{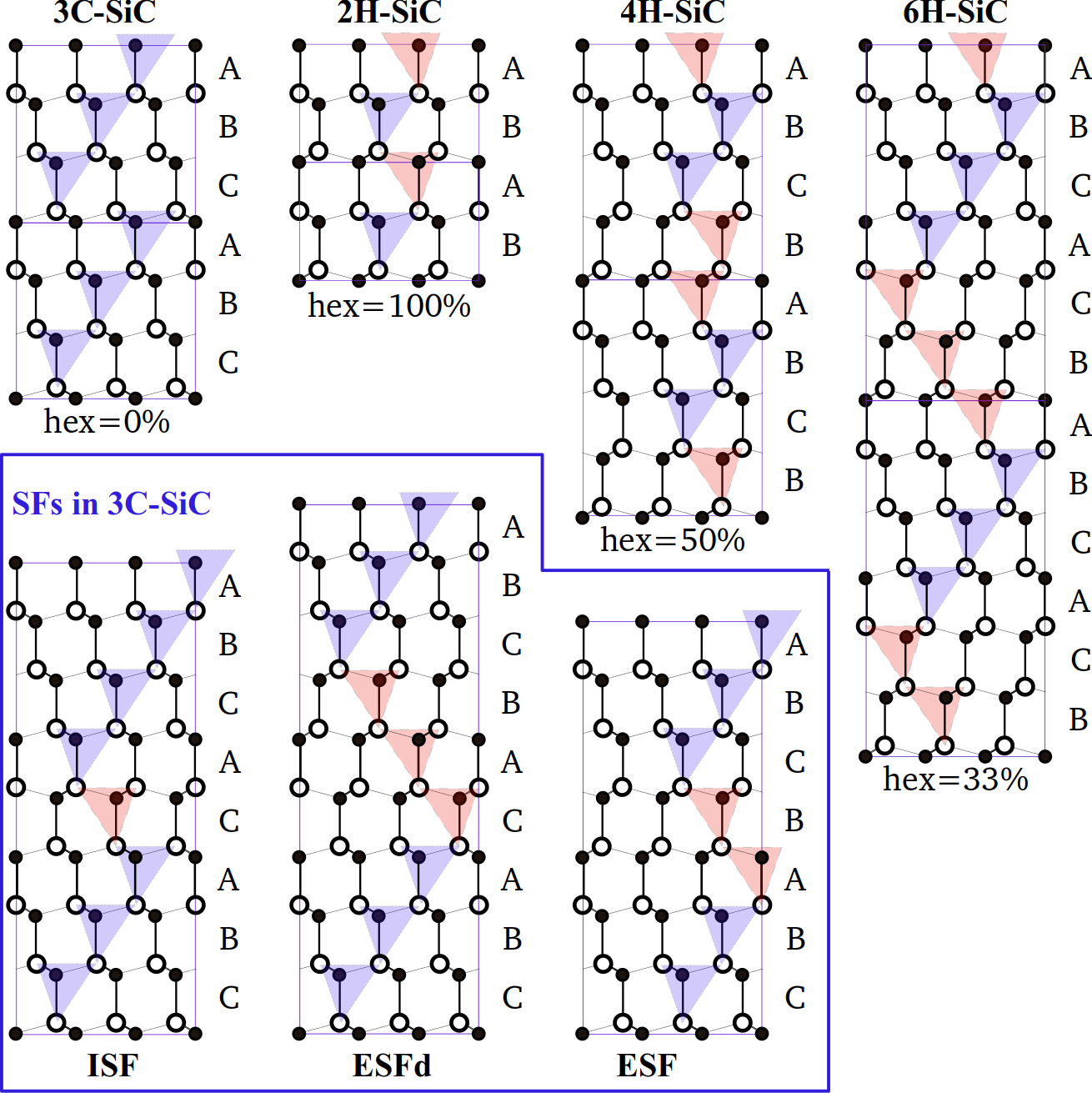}
\caption{\label{sym} Tetrahedral stacking sequences of 3C-, 2H-, 4H- and 6H-SiC. The red and blue triangles highlight the twinned or normal tetrahedra and correspond to down or up spin configurations of the SiC layers, according to the axial next-nearest neighbor Ising (ANNNI) model \cite{Fisher1980InfinitelyModei}.
The inset shows the stacking sequence of 3C-SiC polytype including an intrinsic, extrinsic and double extrinsic stacking fault (labeled ISF, ESF and ESFd respectively). 
}
\end{figure}

Silicon Carbide has become the wide band gap (WBG) semiconductor with the most mature technology \cite{Kimoto2015MaterialAnnealing} and it is finally ready to penetrate the power devices market after more than two decades elapsed as faint promise of next generation power electronics \cite{Chelnokov1997OverviewElectronics,Willander2006SiliconApplications}. Indeed, it is expected that SiC will reach about 10\% of the Si market by 2025 with a compound annual growth rate (CAGR) of about 40\% from 2020 to 2022 \cite{Bhalla2018StatusTechnology}. But, to continue the development of SiC technology and sustain the improvements in efficiency and performance of WBG based devices, research efforts need to be continued even at the level of material physical understanding. In fact, the material maturity process has been quite slow. An evident reason is the intrinsic complexity of this semiconductor compound: it occurs only rarely in nature and it has more than 200 polytypes \cite{N.W.JeppsandT.F.Page1983NoTitle}. If properly understood and controlled, polytypism actually provides an added value. In fact, the most common SiC polytypes (3C-, 4H- and 6H-SiC) cover a range of band-gap from about 2.3 to 3.2 eV and thus they are suitable both for low and high-power devices.\\
Besides the scientific interest, investigating SiC polytypism and understanding its driving force is crucial to correctly predict the energetics of extended defects in SiC, particularly stacking faults (SFs), which are a main concern of this WBG semiconductor since they cause deterioration and eventually failure of the devices after relative long operational time  \cite{Ishida2002InvestigationEpilayers,Nagasawa2008FabricationDefects,Eriksson2011Electrical3C-SiC}.\\
SiC polytypes consist of identical double layers with different stacking sequences, thus generating orders of SiC tetrahedrons with different orientation, as highlighted in blue and red color in Fig.\ref{sym}. Truly, SFs are wrong sequences of the double layers or in other words, they can be seen as inclusions of few layers of a SiC polytype in the perfect layer stacking of another polytype (see inset in Fig.\ref{sym}).  Due to the small-scale energy difference between the stacking sequences of double layers and hence between the different SiC polytypes, as discussed below, perturbations of the ideal stacking sequence during SiC crystal growth are very likely. This is another reason why SFs are so critical in this material. \\
The literature on the thermodynamic stability and polytypism of SiC is abundant, encouraged by the physical and technological interests evidenced above. Still, the predictions on the free energy of SiC polytypes are misleading and the energetic hierarchy reported by different theoretical methods is often inconsistent with some experimental observations. In fact, a paradox concerning SiC polytypes \cite{Heine1991NoTitle,Cheng1990AtomicPolytypes, Fan2014SiliconNanostructures,Zywietz1996InfluenceCarbide} has been often discussed, dealing with the theoretical predictions of hexagonal (6H) SiC as the most stable polytype and of the cubic (3C) one as not stable at any temperature \cite{Cheng1990AtomicPolytypes,Bechstedt1997PolytypismCarbide,Feng2004SiCApplications}. Contrary, experiments have shown that the cubic (3C) structure does grow in preference to all others and only at very high temperatures hexagonal phases, i.e. 4H and 6H polytypes, have been observed to prevail \cite{Boulle2010QuantitativeCrystals,Kanaya1991ControlledDiffraction,Yakimova2000PolytypeBoules,Kado2013High-SpeedMelt,Kusunoki2014Top-seededTechnique,Heine1991NoTitle,Bechstedt1997PolytypismCarbide}. Thereby,  SiC polytype stability at different temperatures remains unclear. Different arguments has been proposed over the years to explain polytypism and polytypic transformation in SiC \cite{Bechstedt1997PolytypismCarbide,Cheng1988Inter-layerPolytypes,Heine1991NoTitle,Ching2006TheSiC,Zywietz1996InfluenceCarbide,Bernstein2005Tight-bindingCarbide,Lindefelt2003StackingPo}, including  the motion of partial dislocations \cite{Pirouz1993PolytypicTEM,Pirouz1997PolytypicSiC,Boulle2010QuantitativeCrystals,Boulle2013PolytypicSimulations} and impurity effects \cite{Heine1991NoTitle} on crystal growth. Nonetheless, inconsistencies between theory and experiments still remain. 

Recently it was shown that density functional theory (DFT) calculations \cite{SM} including the van der Waals (vdW) correction do predict the 3C phase to have the lowest free energy at T=0K \cite{Kawanishi2016EffectPolytypes}. This intriguing result points out the importance of considering long-range interactions when comparing different SiC polytypes.

In this Letter we show that, by consistently adding the entropic contribution to the free energy, within a DFT approach that includes vdW corrections, the full T-dependent hierarchy of polytypes is correctly predicted, showing a cross-over between 3C and 6H (or 4H) phases at typical experimental temperatures. As detailed below, the present results allow for better understanding of the physics behind SiC polytypism by simple thermodynamics considerations, highlighting the correlation between hexagonality, vibrational properties and cohesive energy.
Moreover, calculations of the stability of SiC SFs reveal a T-dependent behavior, intimately correlated to the polytypic stability, and they provide insight into optimal growth temperatures for lowering the density of such defects.

\begin{figure}[!t]
  \centering
    \includegraphics[width=0.48\textwidth]{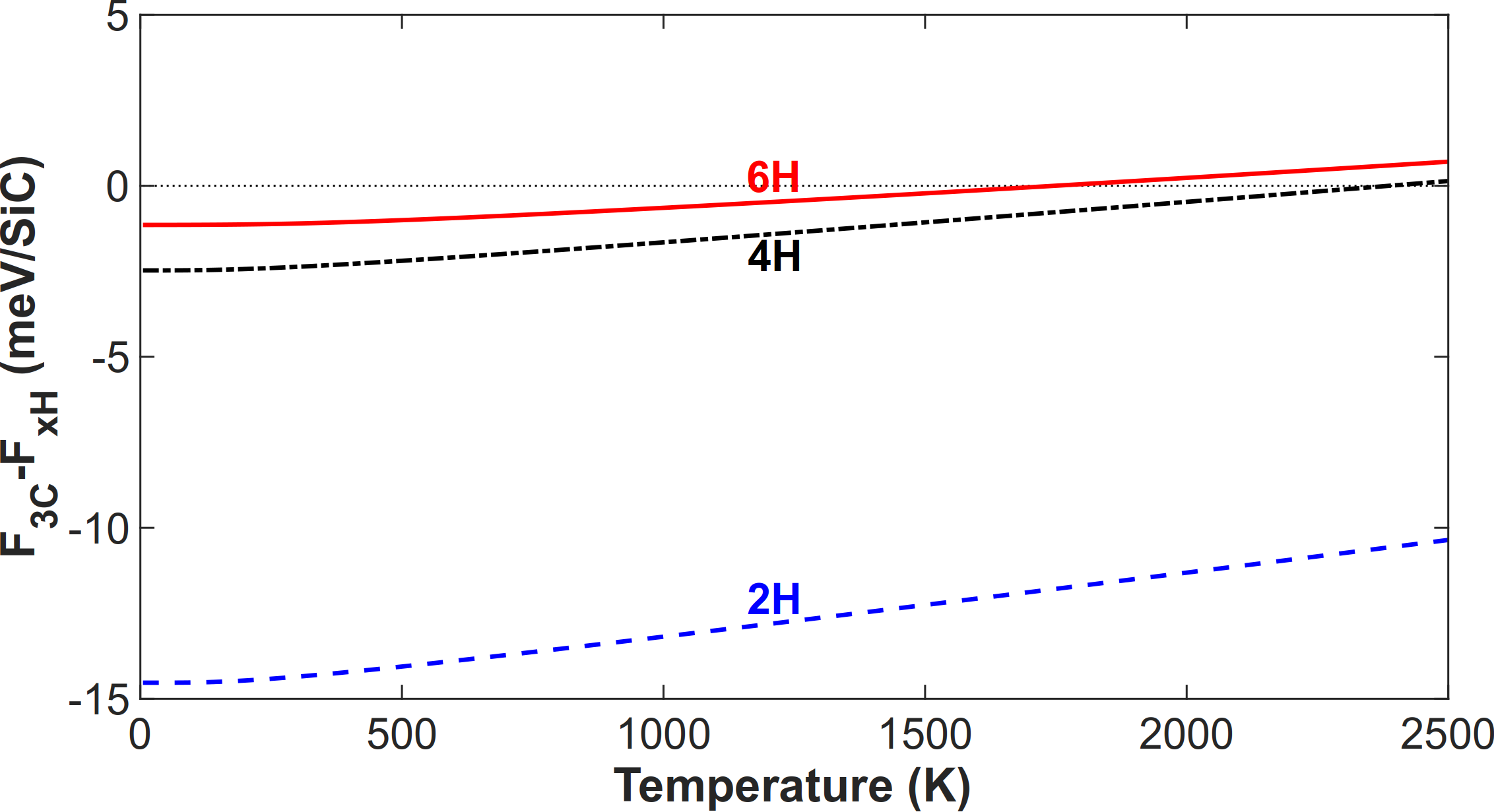}
\caption{\label{HelmohotzF} Difference between the Helmholtz free energy of 3C-SiC with respect to the value of 6H, 4H and 2H polytypes.}
\end{figure}

In Table \ref{tab:table1} the lattice constants of 3C-, 6H-, 4H- and 2H-SiC are reported, both calculated at the generalized gradient approximation (GGA) level \cite{Perdew1996GeneralizedSimple} and with the semiempirical Grimme's method, which employs GGA-type density functional constructed with a long-range dispersion correction \cite{Barone2009RoleCases,Grimme2006SemiempiricalCorrection} and accounting for vdW interactions.
\begin{table}[!t]
\caption{\label{tab:table1}%
Lattice constants (\text{\AA}), total energy $\Delta E_T$ (relative to 3C-SiC), hexagonality and heat of formation $\Delta H_f$ (meV/SiC) of 3C-, 6H-, 4H- and 2H-SiC.}
\begin{ruledtabular}
\begin{tabular}{ l  c c c}
\textrm{Polytype   \hspace{15pt} hex} & 
\textrm{a,c} &
\textrm{$\Delta E_T$}&
\textrm{$\Delta H_f$} \\
\colrule
\textbf{3C-SiC} \hspace{18pt} 0\%\\
GGA & 4.377 & 0 & -402\\
 GGA(vdW) & 4.352 & 0 & -785\\
 exp\cite{o1960silicon, Kleykamp1998GibbsModifications,Greenberg1970TheCalorimetry,2014CRCData}  & 4.3596  &  & -650, -758, -771\\
\colrule

\textbf{6H-SiC} \hspace{18pt} 33\%\\ 
GGA & 3.093,15.178 & -1.7 & -404 \\
 GGA(vdW) & 3.075,15.105 & 1.4 & -784\\
 exp\cite{o1960silicon, Kleykamp1998GibbsModifications,Greenberg1970TheCalorimetry,Tairov1983ProgressCrystals} & 3.080,15.117 & & -676, -747, -771 \\
\colrule

\textbf{4H-SiC} \hspace{18pt} 50\%
\\GGA & 3.092,10.123 & -1.8 & -404\\
  GGA(vdW) & 3.074,10.079 & 2.9  & -783\\
 exp\cite{Zemann1965iCrystalWyckoff,Tairov1983ProgressCrystals,2014CRCData} & 3.073,10.053 & & -650, -689\\
\colrule

\textbf{2H-SiC} \hspace{18pt} 100\%\\
 GGA & 3.090,5.072 & 5.8 & -396\\
  GGA(vdW) & 3.072,5.056 & 15.1 & -770\\
 exp\cite{Schulz1979STRUCTUREZnO} & 3.079,5.053 
%
\end{tabular}
\end{ruledtabular}
\end{table}
It is evident that the agreement between theoretical and experimental lattice parameters improves considerably for vdW-corrected DFT simulations. These improvements are even more appreciable by looking at the heat of formation ($\Delta H_f$) in Table \ref{tab:table1}, which is severely underestimated by GGA but in very good agreement with experiments in the case of vdW-corrected simulations.
Note that, not only the magnitude of the heat of formation but also the order of the values calculated for the different polytypes changes whether or not the simulations include the vdW correction. This is very clear looking at the total energy of the different polytypes ($\Delta E_T$) in Table \ref{tab:table1}, calculated as a relative value with respect to the total energy of 3C-SiC. While $\Delta E_T$ values calculated by GGA are all negative except for 2H-SiC, thus predicting 3C-SiC as the least stable polytype after the 2H-SiC, the vdW-corrected simulations give all positive $\Delta E_T$. Hence, 3C-SiC turns out to be the most stable polytype. This is well in agreement both with experimental evidences inferring 3C-SiC as the most stable SiC structure in the nuclear stage \cite{Tairov1983ProgressCrystals} and with a recent theoretical work \cite{Kawanishi2016EffectPolytypes}. 
Nevertheless, the predicted energetic hierarchy at T=0K is still not sufficient to understand the competition in stability of SiC polytypes at higher temperature, as probed experimentally \cite{Boulle2010QuantitativeCrystals,Boulle2013PolytypicSimulations,Kanaya1991ControlledDiffraction,Yakimova2000PolytypeBoules,Kado2013High-SpeedMelt,Kusunoki2014Top-seededTechnique}. Thus, the variation of the entropic contributions with temperature for the different polytypes becomes crucial. This has been included in Fig.\ref{HelmohotzF}, where the Helmholtz free energy for the SiC polytypes is plotted as a difference between the values of the 2H, 4H and 6H polytypes and that of 3C-SiC. \\
The typical expression of the Helmholtz free energy $F(T)=U-TS$, with $U$ the internal energy and $TS$ the product of temperature and entropy, can be also reformulated as $F(T)=U_0+U_{vib}-TS$, where the internal energy $U$ is split into the vibrational internal energy ($U_{vib}$) and static internal energy ($U_0$), with the latter corresponding to the total DFT energy of the SiC polytypes at their GGA-vdW equilibrium geometry. Also the other terms of the Helmholtz free energy can be conveniently calculated by DFT \cite{Baroni2009ThermalPhonons}. In fact, one can additionally formulate $F(T)$ as a sum of the electronic and vibrational contributions, thus $F(T)=F_{el}(T)+F_{vib}(T)$. The former term $F_{el}(T)$ can be reasonably approximated by its zero-temperature limit $(U_0)$ \cite{Bechstedt1997PolytypismCarbide}, by neglecting the electronic entropy; the vibrational contribution, which then correspond to $F_{vib}(T)=U_{vib}-TS$, can be calculated by the quasi-harmonic approximation as \cite{Born1955DynamicalLattices,Feng2004SiCApplications}:
\begin{equation}
\nonumber 
F_{vib}(T)=\int _{0}^{\infty}g(\omega)  \left [  \frac{\hbar\omega}{2 K_B T} + \ln \left ( 1-e^{  \frac{-\hbar\omega}{K_B T } } \right ) \right ] d\omega \thinspace;
\end{equation}
where $\omega$ is the phonon frequency and $g(\omega)$ is the phonon density of states. In our simulations, $\omega$ and $g(\omega)$ are calculated in the framework of the density functional perturbation theory (DFPT) \cite{Baroni1987Greens-functionSolids,Baroni2009ThermalPhonons}.\\
The values at T=0K of the three curves plotted in Fig.\ref{HelmohotzF} reveal that the zero-point internal energy (ZPE), which is the main contribution to $F_{vib}$ in the low temperature range, only slightly affects the static internal energy (cf. $\Delta E_T$ in Table \ref{tab:table1}) and does not change the energetic hierarchy of the SiC polytypes. But for temperatures above 500K the vibrational contribution ($F_{vib}$) becomes considerable and the Helmohltz free energies of the hexagonal polytypes get closer to the cubic one. Particularly, at temperature of about 1750K the difference between the free energies $(F_{3C}-F_{6H})$ crosses the 0 energy line, meaning that 6H polytype becomes thermodynamically more stable than 3C. The energy crossing between 4H- and 3C-SiC is predicted a bit higher in temperature, at about 2400K. Contrary, 2H- never becomes more stable than 3C-SiC in the temperature range considered, albeit their free energies get closer at higher temperatures. The comparison between hexagonal polytypes reveals that 2H-SiC is the least thermodynamically stable structure: the difference of its free energy and that of 4H or 6H polytype marginally decreases with the temperature. Contrary, the free energies of 6H and 4H polytypes get very close at temperatures around 2500K. Our T-dependent diagram of the polytypic stability plotted in Fig.\ref{HelmohotzF} is in excellent agreement with several experimental evidences such as the preferential growth at temperatures below $\sim$1850K of the 3C polytype over all others \cite{Heine1991NoTitle,Bechstedt1997PolytypismCarbide,Cheng1988Inter-layerPolytypes}, the higher stability of 4H and 6H polytypes at higher temperatures \cite{Boulle2010QuantitativeCrystals,Boulle2013PolytypicSimulations,Kanaya1991ControlledDiffraction,Yakimova2000PolytypeBoules,Kado2013High-SpeedMelt,Kusunoki2014Top-seededTechnique}, and the rare appearance of 2H-SiC \cite{Pirouz1997PolytypicSiC,Imade2009Liquid-phaseMelt}.  

\begin{figure}[t]
  \centering
    \includegraphics[width=0.48\textwidth]{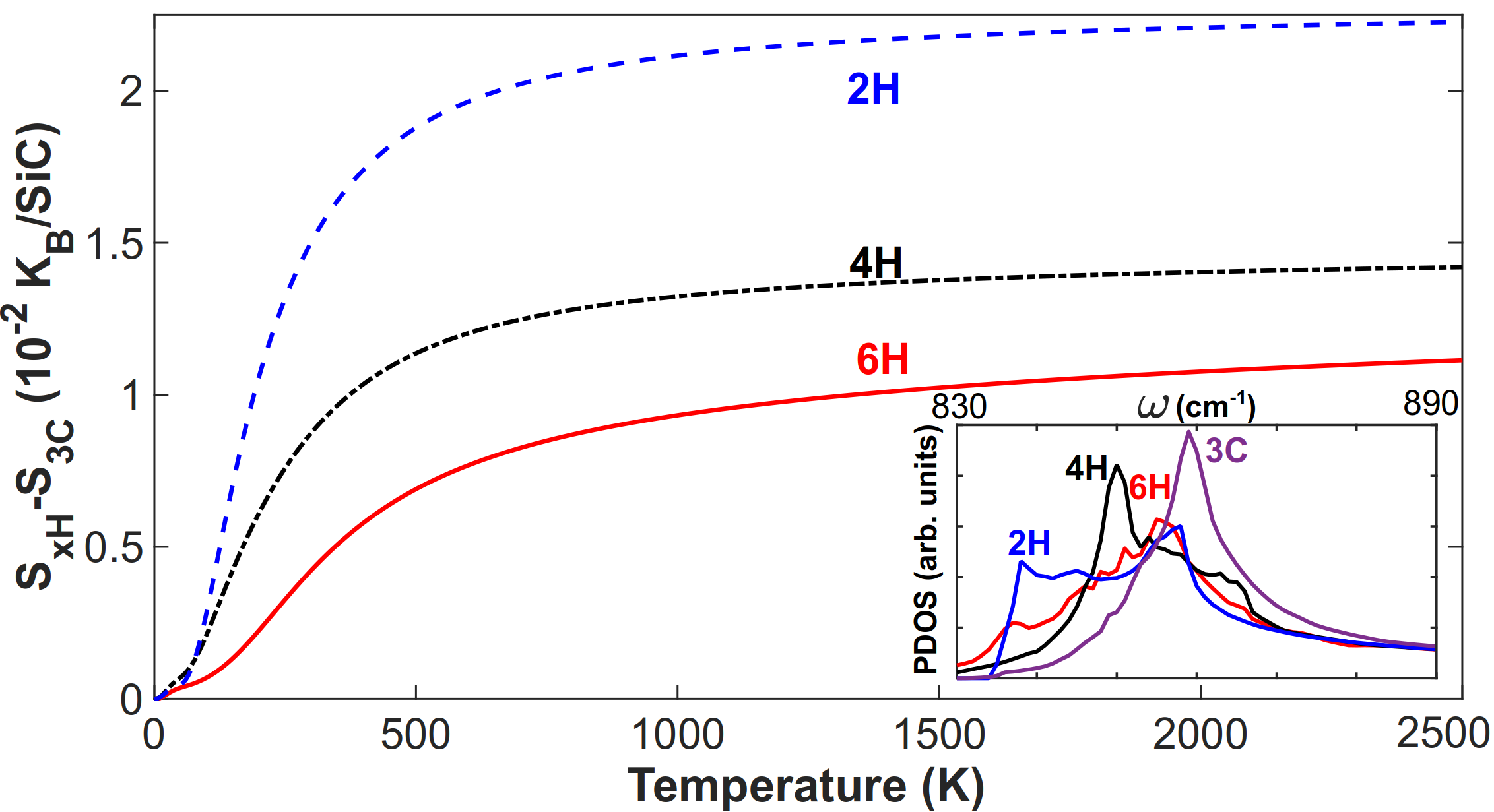}
\caption{\label{FigEntropy} Difference between the entropy of \emph{x}H and 3C politypes. The inset shows the PDOS in the region near the LO band.}
\end{figure}

The correlation between hexagonality and the observed trends in the Helmholtz free energy is elucidated by the calculated entropy, $S=-(\delta F_{vib}/\delta T)$, reported in Fig.\ref{FigEntropy}.
An opposite hierarchy of the entropy with respect to the cohesive energy at all temperatures is found. This is further supported by the general decreasing trend of phonon frequencies with hexagonality, which is evident in the phonon density of states (PDOS) plotted for the different polytypes in the region of the longitudinal optical (LO) branch \cite{SM} in the inset of Fig.\ref{FigEntropy}. The shift of the phonon frequencies is associated to a different strength of the interactions between hexagonally and cubically stacked layers, thus a correspondence between the lower hexagonality and the higher cohesion of the structure is evident.
The difference in the free energy at T=0 between 3C- and 2H-SiC is so large compared to their entropy difference that 2H remains less stable even at high temperature. On the contrary the much smaller difference between the static energy  of 3C and 6H (or 4H) is overcompensated by the larger entropy contribution of the latter.
This provides an intuitive picture to understand SiC polytypism: cubic SiC polytype have higher cohesive energy, higher stiffness and lower entropy; contrary, hexagonal (2H) polytype has the lowest cohesive energy, lower stiffness but a higher entropy; in between, the trends of the other two hexagonal polytypes investigated follow their percentage of hexagonality, with their higher entropy, as compared to 3C-SiC, leading to changes of the energetic hierarchy with the temperature.

\begin{figure}[!t]
  \centering
    \includegraphics[width=0.48\textwidth]{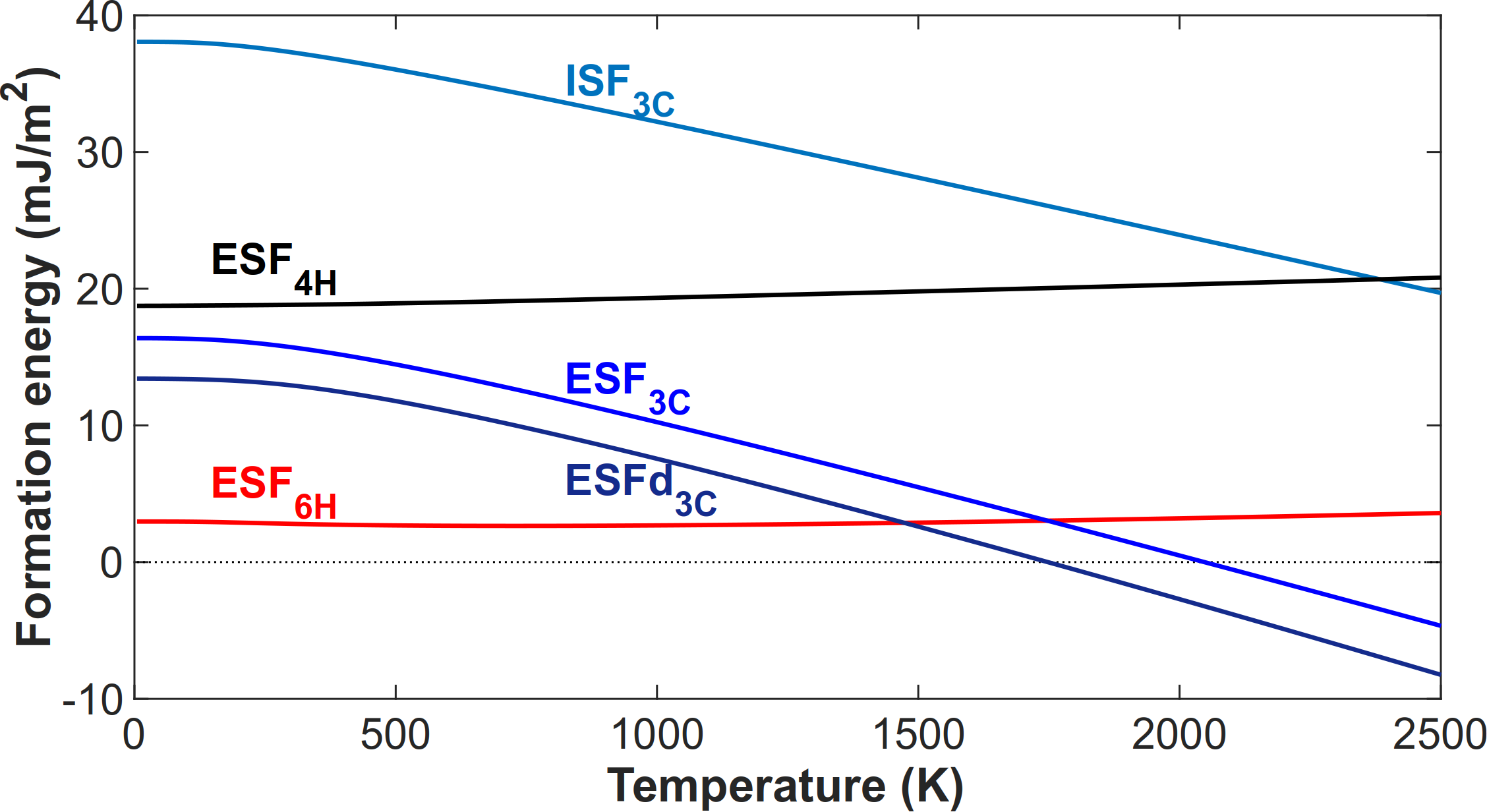}
\caption{\label{defF} Formation energy of ISF in 6H- and 4H-SiC, and of ISF, ESF and ESFd in 3C-SiC.}
\end{figure}

The correct prediction of the free energy of polytypes is a compelling need for the investigation of other essential aspects of SiC and it will be exploited below for studying the SFs stability. 
The energetic cost of any error in the stacking sequence can be estimated by two different approaches: modeling a perturbed layer stacking by supercell structures, such as those illustrated in the inset of Fig.\ref{sym}, and then calculating the total energy of the faulted supercell; alternatively one can calculate the SF energy according to the axial next-nearest neighbor Ising (ANNNI) model \cite{Fisher1980InfinitelyModei}. In fact, the energy of the different polytypes, both pristine or faulted, can be expressed in terms of the interactions between SiC double layers. Accordingly, for the \emph{x}H- (or \emph{x}C-) SiC polytypes the total free energy is:
\begin{equation}
E = E_0 - \frac{1}{x}\displaystyle\sum_{i=1}^n \displaystyle\sum_{j=1}^{\infty}J_j \sigma_i \sigma_{i+j} \thinspace;
\label{eqn:Name}
\end{equation}
where $E_0$ is a common reference energy and $J_j$ are the interaction energies between \emph{i}th-neighbour double layers. The double layers are represented by a pseudospin $\sigma_i$, which can be spin-up or spin-down (with value +1 or -1, respectively) according to the tetrahedron orientation that the layers form: $\sigma_i=+1$ corresponds to a normal orientation (blue color in fig.\ref{sym}) while $\sigma_i=-1$ to a twinned one (red in fig.\ref{sym}) \cite{Cheng1988Inter-layerPolytypes, Pirouz1993PolytypicTEM}. For instance, 4H-SiC is represented by two spin-up and two spin-down, 6H- by three up and three down. In Eq.\ref{eqn:Name}, spin coupling higher than third-order are usually neglected. The static free energies of Table \ref{tab:table1} can be then used to obtain the $J_j$ values from Eq.\ref{eqn:Name} and they allows one to calculate the free energies of faulted structures. Finally, stacking fault energies are estimated as the energy difference between the faulted and the pristine structure \cite{Hong2000StackingCrystals,Boulle2013PolytypicSimulations,Lindefelt2003StackingPo} and they are listed in Table \ref{tab:table2}. 
\begin{table}[!t]
\caption{\label{tab:table2}%
Formation energy (mJ m$^{-2}$) of SFs in 3C, 6H, and 4H-SiC. Values calculated by the ANNNI model (and GGA calculations with/out vdW correction) or obtained by the supercell approach (with vdW) are reported. Experimental values from \cite{Ning1990ExperimentalCrystals,Hong2000StackingCrystals} and other theoretical values from \cite{Umeno2012Ab3C-SiC, Kackell1998StackingStudy,Lindefelt2003StackingPo} are also listed.}
\begin{ruledtabular}
\begin{tabular}{l|ccc|c|c}
        & \multicolumn{3}{c|}{3C-SiC} & 6H-SiC & 4H-SiC \\
        & ISF     & ESF    & ESFd    & ISF    & ISF    \\ \hline
GGA     & 4.35    & -16.7 & -19.5    & 2.84    & 18.23        \\
GGA vdW & 40.70  & 19.6  & 16.85   &    2.77    &   18.35    \\
Superc.& 40.21 & 19.62 & 17.04 & & \\
Exp.    & 34 &  &  &2.9$\pm$0.6&14.7$\pm$2.5 \\
Calc.\cite{Lindefelt2003StackingPo}  & -6.27 &   &   & 3.14 & 18.3\\
Calc.\cite{Umeno2012Ab3C-SiC}  & 10.3 & -7.83 & -11.6 &  &  \\
Calc.\cite{Kackell1998StackingStudy}  & -3.4 & -28 &  &  &  
\end{tabular}
\end{ruledtabular}
\end{table}
We also checked the reliability of the SF energies calculated by the ANNNI model comparing them with the corresponding value obtained by simulating the defected supercells of 3C-SiC illustrated in the inset of Fig.\ref{sym}. Interestingly, the calculated SF energy values are all positive and in excellent  agreement with experimental estimations \cite{Ning1990ExperimentalCrystals,Hong2000StackingCrystals} if the vdW correction is included in the DFT simulations. Contrary, SF energies obtained by bare GGA-DFT are very different, particularly for 3C-SiC. In fact, the formation energy of intrinsic stacking faults (ISFs) in 3C-SiC calculated by GGA is much lower than the corresponding value obtained including the vdW correction. For extrinsic stacking faults, both single (ESF) and double (ESFd), the SF energies turn even into negative values. This is not surprising if one goes through the literature of SFs in 3C-SiC, in which very small or even negative theoretical values of the ESF energy in 3C-SiC are well-accepted (see Table \ref{tab:table2}). Instead, these SF energies are doubtful if compared with experiments \cite{Ning1990ExperimentalCrystals}. Typically, SF energy is experimentally estimated by comparing the measured width of the stacking fault between the two terminating partial dislocations \cite{Ning1990ExperimentalCrystals,Hong2000StackingCrystals} and its expectation by means of the dislocation theory for anisotropic elastic media \cite{Ning1990ExperimentalCrystals,Hong2000StackingCrystals,P.Hirth1982TheoryDislocations}. Accordingly, SFs with negative formation energies should not have finite width, thus in evident contradiction with experiments. \\
Finally, by exploiting the ANNNI model and the T-dependent free energies presented above, we plot in Fig.\ref{defF} the SF formation energies for the 3C-, 6H-, and 4H-SiC as a function of the temperature. Different trends in temperature between hexagonal and cubic polytypes are found: while the formation energies of SF in 6H- and 4H-SiC slightly increase with the temperature, for 3C-SiC the SF energies decrease substantially with the temperature and particularly for the ESFs, they become negative at temperature above 1750K. \\
In conclusion, we have shown that DFT calculations including the vdW correction predict a T-dependent hierarchy of SiC 
polytypes in perfect agreement with the experimental results. 3C-SiC is predicted to have the highest cohesive energy but the lowest entropy. At high temperature, the higher entropic contribution  to the free energy of the hexagonal polytypes stabilizes their structures, with the 6H and 4H-SiC becoming thermodynamically more stable than 3C-SiC. These results demonstrate the key role of the thermodynamics in determining SiC polytypism and contribute to finally reconcile theory and experiments. They are also essential for understanding SF stability in SiC, yielding positive formation-energy values for both ESF and ISF in 3C-SiC that are at variance with previous theoretical results, but in accord with experimental evidences. Moreover, the  formation energy of 3C-SiC SFs is predicted to decrease substantially with temperature, becoming lower than that predicted for 6H-SiC and eventually negative. Importantly, this indicates that too high deposition temperatures should be avoided in order to decrease SF density in 3C-SiC.

\clearpage


\begin{acknowledgements}
Authors acknowledge EU for founding the CHALLENGE project (3C-SiC Hetero-epitaxiALLy grown on silicon compliancE substrates and 3C-SiC substrates for sustaiNable wide-band-Gap powEr devices) within the EU's H2020 framework programme for research and innovation under grant agreement n. 720827.
\end{acknowledgements}
\bibliography{general}

\begin{thebibliography}{48}%
\makeatletter
\providecommand \@ifxundefined [1]{%
 \@ifx{#1\undefined}
}%
\providecommand \@ifnum [1]{%
 \ifnum #1\expandafter \@firstoftwo
 \else \expandafter \@secondoftwo
 \fi
}%
\providecommand \@ifx [1]{%
 \ifx #1\expandafter \@firstoftwo
 \else \expandafter \@secondoftwo
 \fi
}%
\providecommand \natexlab [1]{#1}%
\providecommand \enquote  [1]{``#1''}%
\providecommand \bibnamefont  [1]{#1}%
\providecommand \bibfnamefont [1]{#1}%
\providecommand \citenamefont [1]{#1}%
\providecommand \href@noop [0]{\@secondoftwo}%
\providecommand \href [0]{\begingroup \@sanitize@url \@href}%
\providecommand \@href[1]{\@@startlink{#1}\@@href}%
\providecommand \@@href[1]{\endgroup#1\@@endlink}%
\providecommand \@sanitize@url [0]{\catcode `\\12\catcode `\$12\catcode
  `\&12\catcode `\#12\catcode `\^12\catcode `\_12\catcode `\%12\relax}%
\providecommand \@@startlink[1]{}%
\providecommand \@@endlink[0]{}%
\providecommand \url  [0]{\begingroup\@sanitize@url \@url }%
\providecommand \@url [1]{\endgroup\@href {#1}{\urlprefix }}%
\providecommand \urlprefix  [0]{URL }%
\providecommand \Eprint [0]{\href }%
\providecommand \doibase [0]{http://dx.doi.org/}%
\providecommand \selectlanguage [0]{\@gobble}%
\providecommand \bibinfo  [0]{\@secondoftwo}%
\providecommand \bibfield  [0]{\@secondoftwo}%
\providecommand \translation [1]{[#1]}%
\providecommand \BibitemOpen [0]{}%
\providecommand \bibitemStop [0]{}%
\providecommand \bibitemNoStop [0]{.\EOS\space}%
\providecommand \EOS [0]{\spacefactor3000\relax}%
\providecommand \BibitemShut  [1]{\csname bibitem#1\endcsname}%
\let\auto@bib@innerbib\@empty
\bibitem [{\citenamefont {Kimoto}(2015)}]{Kimoto2015MaterialAnnealing}%
  \BibitemOpen
  \bibfield  {author} {\bibinfo {author} {\bibfnamefont {T.}~\bibnamefont
  {Kimoto}},\ }\href {\doibase 10.7567/JJAP.54.040103} {\bibfield  {journal}
  {\bibinfo  {journal} {Japanese Journal of Applied Physics}\ }\textbf
  {\bibinfo {volume} {54}},\ \bibinfo {pages} {040103} (\bibinfo {year}
  {2015})}\BibitemShut {NoStop}%
\bibitem [{\citenamefont {Chelnokov}\ \emph {et~al.}(1997)\citenamefont
  {Chelnokov}, \citenamefont {Syrkin},\ and\ \citenamefont
  {Dmitriev}}]{Chelnokov1997OverviewElectronics}%
  \BibitemOpen
  \bibfield  {author} {\bibinfo {author} {\bibfnamefont {V.}~\bibnamefont
  {Chelnokov}}, \bibinfo {author} {\bibfnamefont {A.}~\bibnamefont {Syrkin}}, \
  and\ \bibinfo {author} {\bibfnamefont {V.}~\bibnamefont {Dmitriev}},\ }\href
  {\doibase 10.1016/S0925-9635(97)00120-9} {\bibfield  {journal} {\bibinfo
  {journal} {Diamond and Related Materials}\ }\textbf {\bibinfo {volume} {6}},\
  \bibinfo {pages} {1480} (\bibinfo {year} {1997})}\BibitemShut {NoStop}%
\bibitem [{\citenamefont {Willander}\ \emph {et~al.}(2006)\citenamefont
  {Willander}, \citenamefont {Friesel}, \citenamefont {Wahab},\ and\
  \citenamefont {Straumal}}]{Willander2006SiliconApplications}%
  \BibitemOpen
  \bibfield  {author} {\bibinfo {author} {\bibfnamefont {M.}~\bibnamefont
  {Willander}}, \bibinfo {author} {\bibfnamefont {M.}~\bibnamefont {Friesel}},
  \bibinfo {author} {\bibfnamefont {Q.~U.}\ \bibnamefont {Wahab}}, \ and\
  \bibinfo {author} {\bibfnamefont {B.}~\bibnamefont {Straumal}},\ }\href
  {\doibase 10.1007/s10854-005-5137-4} {\bibfield  {journal} {\bibinfo
  {journal} {Journal of Materials Science: Materials in Electronics}\ }\textbf
  {\bibinfo {volume} {17}},\ \bibinfo {pages} {1} (\bibinfo {year}
  {2006})}\BibitemShut {NoStop}%
\bibitem [{\citenamefont {Bhalla}(2018)}]{Bhalla2018StatusTechnology}%
  \BibitemOpen
  \bibfield  {author} {\bibinfo {author} {\bibfnamefont {A.}~\bibnamefont
  {Bhalla}},\ }in\ \href {\doibase 10.5772/intechopen.76061} {\emph {\bibinfo
  {booktitle} {Disruptive Wide Bandgap Semiconductors, Related Technologies,
  and Their Applications}}}\ (\bibinfo  {publisher} {InTech},\ \bibinfo {year}
  {2018})\BibitemShut {NoStop}%
\bibitem [{\citenamefont {{N.W. Jepps and T.F.
  Page}}(1983)}]{N.W.JeppsandT.F.Page1983NoTitle}%
  \BibitemOpen
  \bibfield  {author} {\bibinfo {author} {\bibnamefont {{N.W. Jepps and T.F.
  Page}}},\ }\href@noop {} {\bibfield  {journal} {\bibinfo  {journal} {Prog.
  Cryst. Growth Charact.}\ }\textbf {\bibinfo {volume} {7}},\ \bibinfo {pages}
  {259} (\bibinfo {year} {1983})}\BibitemShut {NoStop}%
\bibitem [{\citenamefont {Ishida}\ \emph {et~al.}(2002)\citenamefont {Ishida},
  \citenamefont {Kushibe}, \citenamefont {Takahashi}, \citenamefont {Okumura},\
  and\ \citenamefont {Yoshida}}]{Ishida2002InvestigationEpilayers}%
  \BibitemOpen
  \bibfield  {author} {\bibinfo {author} {\bibfnamefont {Y.}~\bibnamefont
  {Ishida}}, \bibinfo {author} {\bibfnamefont {M.}~\bibnamefont {Kushibe}},
  \bibinfo {author} {\bibfnamefont {T.}~\bibnamefont {Takahashi}}, \bibinfo
  {author} {\bibfnamefont {H.}~\bibnamefont {Okumura}}, \ and\ \bibinfo
  {author} {\bibfnamefont {S.}~\bibnamefont {Yoshida}},\ }\href {\doibase
  10.4028/www.scientific.net/MSF.389-393.459} {\bibfield  {journal} {\bibinfo
  {journal} {Materials Science Forum}\ }\textbf {\bibinfo {volume} {389-393}},\
  \bibinfo {pages} {459} (\bibinfo {year} {2002})}\BibitemShut {NoStop}%
\bibitem [{\citenamefont {Nagasawa}\ \emph {et~al.}(2008)\citenamefont
  {Nagasawa}, \citenamefont {Abe}, \citenamefont {Yagi}, \citenamefont
  {Kawahara},\ and\ \citenamefont {Hatta}}]{Nagasawa2008FabricationDefects}%
  \BibitemOpen
  \bibfield  {author} {\bibinfo {author} {\bibfnamefont {H.}~\bibnamefont
  {Nagasawa}}, \bibinfo {author} {\bibfnamefont {M.}~\bibnamefont {Abe}},
  \bibinfo {author} {\bibfnamefont {K.}~\bibnamefont {Yagi}}, \bibinfo {author}
  {\bibfnamefont {T.}~\bibnamefont {Kawahara}}, \ and\ \bibinfo {author}
  {\bibfnamefont {N.}~\bibnamefont {Hatta}},\ }\href {\doibase
  10.1002/pssb.200844053} {\bibfield  {journal} {\bibinfo  {journal} {Physica
  Status Solidi (B) Basic Research}\ }\textbf {\bibinfo {volume} {245}},\
  \bibinfo {pages} {1272} (\bibinfo {year} {2008})}\BibitemShut {NoStop}%
\bibitem [{\citenamefont {Eriksson}\ \emph {et~al.}(2011)\citenamefont
  {Eriksson}, \citenamefont {Roccaforte}, \citenamefont {Weng}, \citenamefont
  {Giannazzo}, \citenamefont {Lorenzzi},\ and\ \citenamefont
  {Raineri}}]{Eriksson2011Electrical3C-SiC}%
  \BibitemOpen
  \bibfield  {author} {\bibinfo {author} {\bibfnamefont {J.}~\bibnamefont
  {Eriksson}}, \bibinfo {author} {\bibfnamefont {F.}~\bibnamefont
  {Roccaforte}}, \bibinfo {author} {\bibfnamefont {M.~H.}\ \bibnamefont
  {Weng}}, \bibinfo {author} {\bibfnamefont {F.}~\bibnamefont {Giannazzo}},
  \bibinfo {author} {\bibfnamefont {J.}~\bibnamefont {Lorenzzi}}, \ and\
  \bibinfo {author} {\bibfnamefont {V.}~\bibnamefont {Raineri}},\ }\href
  {\doibase 10.4028/www.scientific.net/MSF.679-680.273} {\bibfield  {journal}
  {\bibinfo  {journal} {Materials Science Forum}\ }\textbf {\bibinfo {volume}
  {679-680}},\ \bibinfo {pages} {273} (\bibinfo {year} {2011})}\BibitemShut
  {NoStop}%
\bibitem [{\citenamefont {Fisher}\ and\ \citenamefont
  {Selke}(1980)}]{Fisher1980InfinitelyModei}%
  \BibitemOpen
  \bibfield  {author} {\bibinfo {author} {\bibfnamefont {M.~E.}\ \bibnamefont
  {Fisher}}\ and\ \bibinfo {author} {\bibfnamefont {W.}~\bibnamefont {Selke}},\
  }\href {https://journals.aps.org/prl/pdf/10.1103/PhysRevLett.44.1502} {\emph
  {\bibinfo {title} {{Infinitely Many Commensurate Phases in a Simple Ising
  Modei}}}},\ \bibinfo {type} {Tech. Rep.}\ (\bibinfo {year}
  {1980})\BibitemShut {NoStop}%
\bibitem [{\citenamefont {Heine}\ \emph {et~al.}(1991)\citenamefont {Heine},
  \citenamefont {Cheng},\ and\ \citenamefont {Needs}}]{Heine1991NoTitle}%
  \BibitemOpen
  \bibfield  {author} {\bibinfo {author} {\bibfnamefont {V.}~\bibnamefont
  {Heine}}, \bibinfo {author} {\bibfnamefont {C.}~\bibnamefont {Cheng}}, \ and\
  \bibinfo {author} {\bibfnamefont {R.~J.}\ \bibnamefont {Needs}},\ }\href@noop
  {} {\bibfield  {journal} {\bibinfo  {journal} {J. Am Ceram SOC.}\ }\textbf
  {\bibinfo {volume} {74}},\ \bibinfo {pages} {2630} (\bibinfo {year}
  {1991})}\BibitemShut {NoStop}%
\bibitem [{\citenamefont {Cheng}\ \emph {et~al.}(1990)\citenamefont {Cheng},
  \citenamefont {Heine},\ and\ \citenamefont
  {Needs}}]{Cheng1990AtomicPolytypes}%
  \BibitemOpen
  \bibfield  {author} {\bibinfo {author} {\bibfnamefont {C.}~\bibnamefont
  {Cheng}}, \bibinfo {author} {\bibfnamefont {V.}~\bibnamefont {Heine}}, \ and\
  \bibinfo {author} {\bibfnamefont {R.~J.}\ \bibnamefont {Needs}},\ }\href
  {http://iopscience.iop.org/article/10.1088/0953-8984/2/23/003/pdf} {\emph
  {\bibinfo {title} {Journal of Physics: Condensed Matter J. Phys.: Condens.
  Matter}}},\ \bibinfo {type} {Tech. Rep.}\ (\bibinfo {year}
  {1990})\BibitemShut {NoStop}%
\bibitem [{\citenamefont {Fan}\ and\ \citenamefont
  {Chu}(2014)}]{Fan2014SiliconNanostructures}%
  \BibitemOpen
  \bibfield  {author} {\bibinfo {author} {\bibfnamefont {J.}~\bibnamefont
  {Fan}}\ and\ \bibinfo {author} {\bibfnamefont {P.~K.}\ \bibnamefont {Chu}},\
  }\href {\doibase 10.1007/978-3-319-08726-9} {\emph {\bibinfo {title}
  {{Silicon Carbide Nanostructures}}}}\ (\bibinfo {year} {2014})\BibitemShut
  {NoStop}%
\bibitem [{\citenamefont {Zywietz}\ \emph {et~al.}(1996)\citenamefont
  {Zywietz}, \citenamefont {Karch},\ and\ \citenamefont
  {Bechstedt}}]{Zywietz1996InfluenceCarbide}%
  \BibitemOpen
  \bibfield  {author} {\bibinfo {author} {\bibfnamefont {A.}~\bibnamefont
  {Zywietz}}, \bibinfo {author} {\bibfnamefont {K.}~\bibnamefont {Karch}}, \
  and\ \bibinfo {author} {\bibfnamefont {F.}~\bibnamefont {Bechstedt}},\ }\href
  {\doibase 10.1103/PhysRevB.54.1791} {\bibfield  {journal} {\bibinfo
  {journal} {Physical Review B - Condensed Matter and Materials Physics}\
  }\textbf {\bibinfo {volume} {54}},\ \bibinfo {pages} {1791} (\bibinfo {year}
  {1996})}\BibitemShut {NoStop}%
\bibitem [{\citenamefont {Bechstedt}\ \emph {et~al.}(1997)\citenamefont
  {Bechstedt}, \citenamefont {K{\"{a}}ckell}, \citenamefont {Zywietz},
  \citenamefont {Karch}, \citenamefont {Adolph}, \citenamefont {Tenelsen},\
  and\ \citenamefont {Furthm{\"{u}}ller}}]{Bechstedt1997PolytypismCarbide}%
  \BibitemOpen
  \bibfield  {author} {\bibinfo {author} {\bibfnamefont {F.}~\bibnamefont
  {Bechstedt}}, \bibinfo {author} {\bibfnamefont {P.}~\bibnamefont
  {K{\"{a}}ckell}}, \bibinfo {author} {\bibfnamefont {A.}~\bibnamefont
  {Zywietz}}, \bibinfo {author} {\bibfnamefont {K.}~\bibnamefont {Karch}},
  \bibinfo {author} {\bibfnamefont {B.}~\bibnamefont {Adolph}}, \bibinfo
  {author} {\bibfnamefont {K.}~\bibnamefont {Tenelsen}}, \ and\ \bibinfo
  {author} {\bibfnamefont {J.}~\bibnamefont {Furthm{\"{u}}ller}},\ }\href
  {\doibase 10.1002/1521-3951(199707)202:1<35::AID-PSSB35>3.0.CO;2-8}
  {\bibfield  {journal} {\bibinfo  {journal} {physica status solidi (b)}\
  }\textbf {\bibinfo {volume} {202}},\ \bibinfo {pages} {35} (\bibinfo {year}
  {1997})}\BibitemShut {NoStop}%
\bibitem [{\citenamefont {Feng}(2004)}]{Feng2004SiCApplications}%
  \BibitemOpen
  \bibfield  {author} {\bibinfo {author} {\bibfnamefont {Z.~C.}\ \bibnamefont
  {Feng}},\ }\href@noop {} {\emph {\bibinfo {title} {{SiC power materials :
  devices and applications}}}}\ (\bibinfo  {publisher} {Springer},\ \bibinfo
  {year} {2004})\ p.\ \bibinfo {pages} {450}\BibitemShut {NoStop}%
\bibitem [{\citenamefont {Boulle}\ \emph {et~al.}(2010)\citenamefont {Boulle},
  \citenamefont {Dompoint}, \citenamefont {Galben-Sandulache},\ and\
  \citenamefont {Chaussende}}]{Boulle2010QuantitativeCrystals}%
  \BibitemOpen
  \bibfield  {author} {\bibinfo {author} {\bibfnamefont {A.}~\bibnamefont
  {Boulle}}, \bibinfo {author} {\bibfnamefont {D.}~\bibnamefont {Dompoint}},
  \bibinfo {author} {\bibfnamefont {I.}~\bibnamefont {Galben-Sandulache}}, \
  and\ \bibinfo {author} {\bibfnamefont {D.}~\bibnamefont {Chaussende}},\
  }\href {\doibase 10.1107/S0021889810019412} {\bibfield  {journal} {\bibinfo
  {journal} {Journal of Applied Crystallography}\ }\textbf {\bibinfo {volume}
  {43}},\ \bibinfo {pages} {867} (\bibinfo {year} {2010})}\BibitemShut
  {NoStop}%
\bibitem [{\citenamefont {Kanaya}\ \emph {et~al.}(1991)\citenamefont {Kanaya},
  \citenamefont {Takahashi}, \citenamefont {Fujiwara},\ and\ \citenamefont
  {Moritani}}]{Kanaya1991ControlledDiffraction}%
  \BibitemOpen
  \bibfield  {author} {\bibinfo {author} {\bibfnamefont {M.}~\bibnamefont
  {Kanaya}}, \bibinfo {author} {\bibfnamefont {J.}~\bibnamefont {Takahashi}},
  \bibinfo {author} {\bibfnamefont {Y.}~\bibnamefont {Fujiwara}}, \ and\
  \bibinfo {author} {\bibfnamefont {A.}~\bibnamefont {Moritani}},\ }\href
  {\doibase 10.1063/1.104443} {\bibfield  {journal} {\bibinfo  {journal}
  {Applied Physics Letters}\ }\textbf {\bibinfo {volume} {58}},\ \bibinfo
  {pages} {56} (\bibinfo {year} {1991})}\BibitemShut {NoStop}%
\bibitem [{\citenamefont {Yakimova}\ \emph {et~al.}(2000)\citenamefont
  {Yakimova}, \citenamefont {Syv{\"{a}}j{\"{a}}rvi}, \citenamefont {Iakimov},
  \citenamefont {Jacobsson}, \citenamefont {R{\aa}back}, \citenamefont
  {Vehanen},\ and\ \citenamefont {Janz{\'{e}}n}}]{Yakimova2000PolytypeBoules}%
  \BibitemOpen
  \bibfield  {author} {\bibinfo {author} {\bibfnamefont {R.}~\bibnamefont
  {Yakimova}}, \bibinfo {author} {\bibfnamefont {M.}~\bibnamefont
  {Syv{\"{a}}j{\"{a}}rvi}}, \bibinfo {author} {\bibfnamefont {T.}~\bibnamefont
  {Iakimov}}, \bibinfo {author} {\bibfnamefont {H.}~\bibnamefont {Jacobsson}},
  \bibinfo {author} {\bibfnamefont {R.}~\bibnamefont {R{\aa}back}}, \bibinfo
  {author} {\bibfnamefont {A.}~\bibnamefont {Vehanen}}, \ and\ \bibinfo
  {author} {\bibfnamefont {E.}~\bibnamefont {Janz{\'{e}}n}},\ }\href {\doibase
  10.1016/S0022-0248(00)00488-7} {\bibfield  {journal} {\bibinfo  {journal}
  {Journal of Crystal Growth}\ }\textbf {\bibinfo {volume} {217}},\ \bibinfo
  {pages} {255} (\bibinfo {year} {2000})}\BibitemShut {NoStop}%
\bibitem [{\citenamefont {Kado}\ \emph {et~al.}(2013)\citenamefont {Kado} \emph
  {et~al.}}]{Kado2013High-SpeedMelt}%
  \BibitemOpen
  \bibfield  {author} {\bibinfo {author} {\bibfnamefont {M.}~\bibnamefont
  {Kado}} \emph {et~al.},\ }\href {\doibase
  10.4028/www.scientific.net/MSF.740-742.73} {\bibfield  {journal} {\bibinfo
  {journal} {Materials Science Forum}\ }\textbf {\bibinfo {volume} {740-742}},\
  \bibinfo {pages} {73} (\bibinfo {year} {2013})}\BibitemShut {NoStop}%
\bibitem [{\citenamefont {Kusunoki}\ \emph {et~al.}(2014)\citenamefont
  {Kusunoki}, \citenamefont {Okada}, \citenamefont {Kamei}, \citenamefont
  {Moriguchi}, \citenamefont {Daikoku}, \citenamefont {Kado}, \citenamefont
  {Sakamoto}, \citenamefont {Bessho},\ and\ \citenamefont
  {Ujihara}}]{Kusunoki2014Top-seededTechnique}%
  \BibitemOpen
  \bibfield  {author} {\bibinfo {author} {\bibfnamefont {K.}~\bibnamefont
  {Kusunoki}}, \bibinfo {author} {\bibfnamefont {N.}~\bibnamefont {Okada}},
  \bibinfo {author} {\bibfnamefont {K.}~\bibnamefont {Kamei}}, \bibinfo
  {author} {\bibfnamefont {K.}~\bibnamefont {Moriguchi}}, \bibinfo {author}
  {\bibfnamefont {H.}~\bibnamefont {Daikoku}}, \bibinfo {author} {\bibfnamefont
  {M.}~\bibnamefont {Kado}}, \bibinfo {author} {\bibfnamefont {H.}~\bibnamefont
  {Sakamoto}}, \bibinfo {author} {\bibfnamefont {T.}~\bibnamefont {Bessho}}, \
  and\ \bibinfo {author} {\bibfnamefont {T.}~\bibnamefont {Ujihara}},\ }\href
  {\doibase 10.1016/j.jcrysgro.2014.03.006} {\bibfield  {journal} {\bibinfo
  {journal} {Journal of Crystal Growth}\ }\textbf {\bibinfo {volume} {395}},\
  \bibinfo {pages} {68} (\bibinfo {year} {2014})}\BibitemShut {NoStop}%
\bibitem [{\citenamefont {Cheng}\ \emph {et~al.}(1988)\citenamefont {Cheng},
  \citenamefont {Needs},\ and\ \citenamefont
  {Heine}}]{Cheng1988Inter-layerPolytypes}%
  \BibitemOpen
  \bibfield  {author} {\bibinfo {author} {\bibfnamefont {C.}~\bibnamefont
  {Cheng}}, \bibinfo {author} {\bibfnamefont {R.~J.}\ \bibnamefont {Needs}}, \
  and\ \bibinfo {author} {\bibfnamefont {V.}~\bibnamefont {Heine}},\ }\href
  {\doibase 10.1088/0022-3719/21/6/012} {\bibfield  {journal} {\bibinfo
  {journal} {Journal of Physics C: Solid State Physics}\ }\textbf {\bibinfo
  {volume} {21}},\ \bibinfo {pages} {1049} (\bibinfo {year}
  {1988})}\BibitemShut {NoStop}%
\bibitem [{\citenamefont {Ching}\ \emph {et~al.}(2006)\citenamefont {Ching},
  \citenamefont {Xu}, \citenamefont {Rulis},\ and\ \citenamefont
  {Ouyang}}]{Ching2006TheSiC}%
  \BibitemOpen
  \bibfield  {author} {\bibinfo {author} {\bibfnamefont {W.~Y.}\ \bibnamefont
  {Ching}}, \bibinfo {author} {\bibfnamefont {Y.~N.}\ \bibnamefont {Xu}},
  \bibinfo {author} {\bibfnamefont {P.}~\bibnamefont {Rulis}}, \ and\ \bibinfo
  {author} {\bibfnamefont {L.}~\bibnamefont {Ouyang}},\ }\href {\doibase
  10.1016/j.msea.2006.01.007} {\bibfield  {journal} {\bibinfo  {journal}
  {Materials Science and Engineering A}\ }\textbf {\bibinfo {volume} {422}},\
  \bibinfo {pages} {147} (\bibinfo {year} {2006})}\BibitemShut {NoStop}%
\bibitem [{\citenamefont {Bernstein}\ \emph {et~al.}(2005)\citenamefont
  {Bernstein}, \citenamefont {Gotsis}, \citenamefont {Papaconstantopoulos},\
  and\ \citenamefont {Mehl}}]{Bernstein2005Tight-bindingCarbide}%
  \BibitemOpen
  \bibfield  {author} {\bibinfo {author} {\bibfnamefont {N.}~\bibnamefont
  {Bernstein}}, \bibinfo {author} {\bibfnamefont {H.~J.}\ \bibnamefont
  {Gotsis}}, \bibinfo {author} {\bibfnamefont {D.~A.}\ \bibnamefont
  {Papaconstantopoulos}}, \ and\ \bibinfo {author} {\bibfnamefont {M.~J.}\
  \bibnamefont {Mehl}},\ }\href {\doibase 10.1103/PhysRevB.71.075203} {\ ,\
  \bibinfo {pages} {1} (\bibinfo {year} {2005})}\BibitemShut {NoStop}%
\bibitem [{\citenamefont {Lindefelt}\ \emph {et~al.}(2003)\citenamefont
  {Lindefelt}, \citenamefont {Iwata}, \citenamefont {{\"{O}}berg},\ and\
  \citenamefont {Briddon}}]{Lindefelt2003StackingPo}%
  \BibitemOpen
  \bibfield  {author} {\bibinfo {author} {\bibfnamefont {U.}~\bibnamefont
  {Lindefelt}}, \bibinfo {author} {\bibfnamefont {H.}~\bibnamefont {Iwata}},
  \bibinfo {author} {\bibfnamefont {S.}~\bibnamefont {{\"{O}}berg}}, \ and\
  \bibinfo {author} {\bibfnamefont {P.~R.}\ \bibnamefont {Briddon}},\ }\href
  {\doibase 10.1103/PhysRevB.67.155204} {\bibfield  {journal} {\bibinfo
  {journal} {Physical Review B}\ }\textbf {\bibinfo {volume} {67}},\ \bibinfo
  {pages} {155204} (\bibinfo {year} {2003})}\BibitemShut {NoStop}%
\bibitem [{\citenamefont {Pirouz}\ and\ \citenamefont
  {Yang}(1993)}]{Pirouz1993PolytypicTEM}%
  \BibitemOpen
  \bibfield  {author} {\bibinfo {author} {\bibfnamefont {P.}~\bibnamefont
  {Pirouz}}\ and\ \bibinfo {author} {\bibfnamefont {J.}~\bibnamefont {Yang}},\
  }\href {\doibase 10.1016/0304-3991(93)90146-O} {\bibfield  {journal}
  {\bibinfo  {journal} {Ultramicroscopy}\ }\textbf {\bibinfo {volume} {51}},\
  \bibinfo {pages} {189} (\bibinfo {year} {1993})}\BibitemShut {NoStop}%
\bibitem [{\citenamefont {Pirouz}(1997)}]{Pirouz1997PolytypicSiC}%
  \BibitemOpen
  \bibfield  {author} {\bibinfo {author} {\bibfnamefont {P.}~\bibnamefont
  {Pirouz}},\ }\href {\doibase 10.4028/www.scientific.net/SSP.56.107}
  {\bibfield  {journal} {\bibinfo  {journal} {Solid State Phenomena}\ }\textbf
  {\bibinfo {volume} {56}},\ \bibinfo {pages} {107} (\bibinfo {year}
  {1997})}\BibitemShut {NoStop}%
\bibitem [{\citenamefont {Boulle}\ \emph {et~al.}(2013)\citenamefont {Boulle},
  \citenamefont {Dompoint}, \citenamefont {Galben-Sandulache},\ and\
  \citenamefont {Chaussende}}]{Boulle2013PolytypicSimulations}%
  \BibitemOpen
  \bibfield  {author} {\bibinfo {author} {\bibfnamefont {A.}~\bibnamefont
  {Boulle}}, \bibinfo {author} {\bibfnamefont {D.}~\bibnamefont {Dompoint}},
  \bibinfo {author} {\bibfnamefont {I.}~\bibnamefont {Galben-Sandulache}}, \
  and\ \bibinfo {author} {\bibfnamefont {D.}~\bibnamefont {Chaussende}},\
  }\href {\doibase 10.1103/PhysRevB.88.024103} {\bibfield  {journal} {\bibinfo
  {journal} {Physical Review B - Condensed Matter and Materials Physics}\
  }\textbf {\bibinfo {volume} {88}} (\bibinfo {year} {2013}),\
  10.1103/PhysRevB.88.024103}\BibitemShut {NoStop}%
\bibitem [{SM()}]{SM}%
  \BibitemOpen
  \href@noop {} {}\bibinfo {note} {See Supplemental Material for details on the
  theoretical method and for the complete PDOS plot.}\BibitemShut {Stop}%
\bibitem [{\citenamefont {Kawanishi}\ and\ \citenamefont
  {Mizoguchi}(2016)}]{Kawanishi2016EffectPolytypes}%
  \BibitemOpen
  \bibfield  {author} {\bibinfo {author} {\bibfnamefont {S.}~\bibnamefont
  {Kawanishi}}\ and\ \bibinfo {author} {\bibfnamefont {T.}~\bibnamefont
  {Mizoguchi}},\ }\href {\doibase 10.1063/1.4948329} {\bibfield  {journal}
  {\bibinfo  {journal} {Journal of Applied Physics}\ }\textbf {\bibinfo
  {volume} {119}},\ \bibinfo {pages} {175101} (\bibinfo {year}
  {2016})}\BibitemShut {NoStop}%
\bibitem [{\citenamefont {Perdew}\ \emph {et~al.}(1996)\citenamefont {Perdew},
  \citenamefont {Burke},\ and\ \citenamefont
  {Ernzerhof}}]{Perdew1996GeneralizedSimple}%
  \BibitemOpen
  \bibfield  {author} {\bibinfo {author} {\bibfnamefont {J.~P.}\ \bibnamefont
  {Perdew}}, \bibinfo {author} {\bibfnamefont {K.}~\bibnamefont {Burke}}, \
  and\ \bibinfo {author} {\bibfnamefont {M.}~\bibnamefont {Ernzerhof}},\ }\href
  {\doibase 10.1103/PhysRevLett.77.3865} {\bibfield  {journal} {\bibinfo
  {journal} {Physical Review Letters}\ }\textbf {\bibinfo {volume} {77}},\
  \bibinfo {pages} {3865} (\bibinfo {year} {1996})}\BibitemShut {NoStop}%
\bibitem [{\citenamefont {Barone}\ \emph {et~al.}(2009)\citenamefont {Barone},
  \citenamefont {Casarin}, \citenamefont {Forrer}, \citenamefont {Pavone},
  \citenamefont {Sambi},\ and\ \citenamefont
  {Vittadini}}]{Barone2009RoleCases}%
  \BibitemOpen
  \bibfield  {author} {\bibinfo {author} {\bibfnamefont {V.}~\bibnamefont
  {Barone}}, \bibinfo {author} {\bibfnamefont {M.}~\bibnamefont {Casarin}},
  \bibinfo {author} {\bibfnamefont {D.}~\bibnamefont {Forrer}}, \bibinfo
  {author} {\bibfnamefont {M.}~\bibnamefont {Pavone}}, \bibinfo {author}
  {\bibfnamefont {M.}~\bibnamefont {Sambi}}, \ and\ \bibinfo {author}
  {\bibfnamefont {A.}~\bibnamefont {Vittadini}},\ }\href {\doibase
  10.1002/jcc.21112} {\bibfield  {journal} {\bibinfo  {journal} {Journal of
  Computational Chemistry}\ }\textbf {\bibinfo {volume} {30}},\ \bibinfo
  {pages} {934} (\bibinfo {year} {2009})}\BibitemShut {NoStop}%
\bibitem [{\citenamefont {Grimme}(2006)}]{Grimme2006SemiempiricalCorrection}%
  \BibitemOpen
  \bibfield  {author} {\bibinfo {author} {\bibfnamefont {S.}~\bibnamefont
  {Grimme}},\ }\href {\doibase 10.1002/jcc.20495} {\bibfield  {journal}
  {\bibinfo  {journal} {Journal of Computational Chemistry}\ }\textbf {\bibinfo
  {volume} {27}},\ \bibinfo {pages} {1787} (\bibinfo {year}
  {2006})}\BibitemShut {NoStop}%
\bibitem [{\citenamefont {O'Connor}\ \emph {et~al.}(1960)\citenamefont
  {O'Connor}, \citenamefont {Smiltens},\ and\ \citenamefont
  {Directorate}}]{o1960silicon}%
  \BibitemOpen
  \bibfield  {author} {\bibinfo {author} {\bibfnamefont {J.~R.}\ \bibnamefont
  {O'Connor}}, \bibinfo {author} {\bibfnamefont {J.}~\bibnamefont {Smiltens}},
  \ and\ \bibinfo {author} {\bibnamefont {Directorate}},\ }\href
  {https://books.google.it/books?id=huANAQAAIAAJ} {\emph {\bibinfo {title}
  {{Silicon Carbide, a High Temperature Semiconductor: Proceedings}}}}\
  (\bibinfo  {publisher} {Symposium Publications Division, Pergamon Press},\
  \bibinfo {year} {1960})\BibitemShut {NoStop}%
\bibitem [{\citenamefont {Kleykamp}(1998)}]{Kleykamp1998GibbsModifications}%
  \BibitemOpen
  \bibfield  {author} {\bibinfo {author} {\bibfnamefont {H.}~\bibnamefont
  {Kleykamp}},\ }\href {\doibase 10.1002/bbpc.19981020928} {\bibfield
  {journal} {\bibinfo  {journal} {Berichte der Bunsengesellschaft f{\"{u}}r
  physikalische Chemie}\ }\textbf {\bibinfo {volume} {102}},\ \bibinfo {pages}
  {1231} (\bibinfo {year} {1998})}\BibitemShut {NoStop}%
\bibitem [{\citenamefont {Greenberg}\ \emph {et~al.}(1970)\citenamefont
  {Greenberg}, \citenamefont {Natke},\ and\ \citenamefont
  {Hubbard}}]{Greenberg1970TheCalorimetry}%
  \BibitemOpen
  \bibfield  {author} {\bibinfo {author} {\bibfnamefont {E.}~\bibnamefont
  {Greenberg}}, \bibinfo {author} {\bibfnamefont {C.~A.}\ \bibnamefont
  {Natke}}, \ and\ \bibinfo {author} {\bibfnamefont {W.~N.}\ \bibnamefont
  {Hubbard}},\ }\href {\doibase 10.1016/0021-9614(70)90083-2} {\bibfield
  {journal} {\bibinfo  {journal} {The Journal of Chemical Thermodynamics}\
  }\textbf {\bibinfo {volume} {2}},\ \bibinfo {pages} {193} (\bibinfo {year}
  {1970})}\BibitemShut {NoStop}%
\bibitem [{\citenamefont {Haynes}(2014)}]{2014CRCData}%
  \BibitemOpen
  \bibinfo {editor} {\bibfnamefont {W.~M.}\ \bibnamefont {Haynes}},\ ed.,\
  \href {http://www.hbcpnetbase.com/} {\emph {\bibinfo {title} {{CRC handbook
  of chemistry and physics: a ready-reference book of chemical and physical
  data}}}},\ \bibinfo {edition} {95th}\ ed.\ (\bibinfo  {publisher} {CRC
  Press},\ \bibinfo {address} {Boca Raton},\ \bibinfo {year}
  {2014})\BibitemShut {NoStop}%
\bibitem [{\citenamefont {Tairov}\ and\ \citenamefont
  {Tsvetkov}(1983)}]{Tairov1983ProgressCrystals}%
  \BibitemOpen
  \bibfield  {author} {\bibinfo {author} {\bibfnamefont {Y.}~\bibnamefont
  {Tairov}}\ and\ \bibinfo {author} {\bibfnamefont {V.}~\bibnamefont
  {Tsvetkov}},\ }\href {\doibase 10.1016/0146-3535(83)90031-X} {\bibfield
  {journal} {\bibinfo  {journal} {Progress in Crystal Growth and
  Characterization}\ }\textbf {\bibinfo {volume} {7}},\ \bibinfo {pages} {111}
  (\bibinfo {year} {1983})}\BibitemShut {NoStop}%
\bibitem [{\citenamefont {Zemann}(1965)}]{Zemann1965iCrystalWyckoff}%
  \BibitemOpen
  \bibfield  {author} {\bibinfo {author} {\bibfnamefont {J.}~\bibnamefont
  {Zemann}},\ }\href {\doibase 10.1107/S0365110X65000361} {\bibfield  {journal}
  {\bibinfo  {journal} {Acta Crystallographica}\ }\textbf {\bibinfo {volume}
  {18}},\ \bibinfo {pages} {139} (\bibinfo {year} {1965})}\BibitemShut
  {NoStop}%
\bibitem [{\citenamefont {Schulz}\ and\ \citenamefont
  {Thiemann}(1979)}]{Schulz1979STRUCTUREZnO}%
  \BibitemOpen
  \bibfield  {author} {\bibinfo {author} {\bibfnamefont {H.}~\bibnamefont
  {Schulz}}\ and\ \bibinfo {author} {\bibfnamefont {K.~H.}\ \bibnamefont
  {Thiemann}},\ }\href
  {https://ac.els-cdn.com/0038109879907543/1-s2.0-0038109879907543-main.pdf?_tid=20fbabc8-8900-417a-8692-2e084767e25d&acdnat=1548174058_9f76402c36d7441e047523711d850538}
  {\emph {\bibinfo {title} {Solid State Communications}}},\ \bibinfo {type}
  {Tech. Rep.}\ (\bibinfo {year} {1979})\BibitemShut {NoStop}%
\bibitem [{\citenamefont {Baroni}\ \emph {et~al.}(2009)\citenamefont {Baroni},
  \citenamefont {Giannozzi},\ and\ \citenamefont
  {Isaev}}]{Baroni2009ThermalPhonons}%
  \BibitemOpen
  \bibfield  {author} {\bibinfo {author} {\bibfnamefont {S.}~\bibnamefont
  {Baroni}}, \bibinfo {author} {\bibfnamefont {P.}~\bibnamefont {Giannozzi}}, \
  and\ \bibinfo {author} {\bibfnamefont {E.}~\bibnamefont {Isaev}},\ }\href
  {\doibase 10.2138/rmg.2009.71.3} {\bibfield  {journal} {\bibinfo  {journal}
  {Reviews in Mineralogy {\&} Geochemistry}\ }\textbf {\bibinfo {volume} {71}}
  (\bibinfo {year} {2009}),\ 10.2138/rmg.2009.71.3}\BibitemShut {NoStop}%
\bibitem [{\citenamefont {Born}\ \emph {et~al.}(1955)\citenamefont {Born},
  \citenamefont {Huang},\ and\ \citenamefont
  {Lax}}]{Born1955DynamicalLattices}%
  \BibitemOpen
  \bibfield  {author} {\bibinfo {author} {\bibfnamefont {M.}~\bibnamefont
  {Born}}, \bibinfo {author} {\bibfnamefont {K.}~\bibnamefont {Huang}}, \ and\
  \bibinfo {author} {\bibfnamefont {M.}~\bibnamefont {Lax}},\ }\href {\doibase
  10.1119/1.1934059} {\bibfield  {journal} {\bibinfo  {journal} {American
  Journal of Physics}\ }\textbf {\bibinfo {volume} {23}},\ \bibinfo {pages}
  {474} (\bibinfo {year} {1955})}\BibitemShut {NoStop}%
\bibitem [{\citenamefont {Baroni}\ \emph {et~al.}(1987)\citenamefont {Baroni},
  \citenamefont {Giannozzi},\ and\ \citenamefont
  {Testa}}]{Baroni1987Greens-functionSolids}%
  \BibitemOpen
  \bibfield  {author} {\bibinfo {author} {\bibfnamefont {S.}~\bibnamefont
  {Baroni}}, \bibinfo {author} {\bibfnamefont {P.}~\bibnamefont {Giannozzi}}, \
  and\ \bibinfo {author} {\bibfnamefont {A.}~\bibnamefont {Testa}},\ }\href
  {\doibase 10.1103/PhysRevLett.58.1861} {\bibfield  {journal} {\bibinfo
  {journal} {Physical Review Letters}\ }\textbf {\bibinfo {volume} {58}},\
  \bibinfo {pages} {1861} (\bibinfo {year} {1987})}\BibitemShut {NoStop}%
\bibitem [{\citenamefont {Imade}\ \emph {et~al.}(2009)\citenamefont {Imade}
  \emph {et~al.}}]{Imade2009Liquid-phaseMelt}%
  \BibitemOpen
  \bibfield  {author} {\bibinfo {author} {\bibfnamefont {M.}~\bibnamefont
  {Imade}} \emph {et~al.},\ }\href {\doibase 10.1016/j.matlet.2008.12.009}
  {\bibfield  {journal} {\bibinfo  {journal} {Materials Letters}\ }\textbf
  {\bibinfo {volume} {63}},\ \bibinfo {pages} {649} (\bibinfo {year}
  {2009})}\BibitemShut {NoStop}%
\bibitem [{\citenamefont {Hong}\ \emph {et~al.}(2000)\citenamefont {Hong},
  \citenamefont {Samant},\ and\ \citenamefont
  {Pirouz}}]{Hong2000StackingCrystals}%
  \BibitemOpen
  \bibfield  {author} {\bibinfo {author} {\bibfnamefont {M.~H.}\ \bibnamefont
  {Hong}}, \bibinfo {author} {\bibfnamefont {A.~V.}\ \bibnamefont {Samant}}, \
  and\ \bibinfo {author} {\bibfnamefont {P.}~\bibnamefont {Pirouz}},\ }\href
  {\doibase 10.1080/01418610008212090} {\bibfield  {journal} {\bibinfo
  {journal} {Philosophical Magazine A: Physics of Condensed Matter, Structure,
  Defects and Mechanical Properties}\ }\textbf {\bibinfo {volume} {80}},\
  \bibinfo {pages} {919} (\bibinfo {year} {2000})}\BibitemShut {NoStop}%
\bibitem [{\citenamefont {Ning}\ and\ \citenamefont
  {Ye}(1990)}]{Ning1990ExperimentalCrystals}%
  \BibitemOpen
  \bibfield  {author} {\bibinfo {author} {\bibfnamefont {X.~G.}\ \bibnamefont
  {Ning}}\ and\ \bibinfo {author} {\bibfnamefont {H.~Q.}\ \bibnamefont {Ye}},\
  }\href {\doibase 10.1088/0953-8984/2/50/028} {\bibfield  {journal} {\bibinfo
  {journal} {Journal of Physics: Condensed Matter}\ }\textbf {\bibinfo {volume}
  {2}},\ \bibinfo {pages} {10223} (\bibinfo {year} {1990})}\BibitemShut
  {NoStop}%
\bibitem [{\citenamefont {Umeno}\ \emph {et~al.}(2012)\citenamefont {Umeno},
  \citenamefont {Yagi},\ and\ \citenamefont {Nagasawa}}]{Umeno2012Ab3C-SiC}%
  \BibitemOpen
  \bibfield  {author} {\bibinfo {author} {\bibfnamefont {Y.}~\bibnamefont
  {Umeno}}, \bibinfo {author} {\bibfnamefont {K.}~\bibnamefont {Yagi}}, \ and\
  \bibinfo {author} {\bibfnamefont {H.}~\bibnamefont {Nagasawa}},\ }\href
  {\doibase 10.1002/pssb.201147487} {\bibfield  {journal} {\bibinfo  {journal}
  {physica status solidi (b)}\ }\textbf {\bibinfo {volume} {249}},\ \bibinfo
  {pages} {1229} (\bibinfo {year} {2012})}\BibitemShut {NoStop}%
\bibitem [{\citenamefont {K{\"{a}}ckell}\ \emph {et~al.}(1998)\citenamefont
  {K{\"{a}}ckell}, \citenamefont {Furthm{\"{u}}ller},\ and\ \citenamefont
  {Bechstedt}}]{Kackell1998StackingStudy}%
  \BibitemOpen
  \bibfield  {author} {\bibinfo {author} {\bibfnamefont {P.}~\bibnamefont
  {K{\"{a}}ckell}}, \bibinfo {author} {\bibfnamefont {J.}~\bibnamefont
  {Furthm{\"{u}}ller}}, \ and\ \bibinfo {author} {\bibfnamefont
  {F.}~\bibnamefont {Bechstedt}},\ }\href {\doibase 10.1103/PhysRevB.58.1326}
  {\bibfield  {journal} {\bibinfo  {journal} {Physical Review B - Condensed
  Matter and Materials Physics}\ }\textbf {\bibinfo {volume} {58}},\ \bibinfo
  {pages} {1326} (\bibinfo {year} {1998})}\BibitemShut {NoStop}%
\bibitem [{\citenamefont {P.~Hirth}\ and\ \citenamefont
  {Lothe}(1982)}]{P.Hirth1982TheoryDislocations}%
  \BibitemOpen
  \bibfield  {author} {\bibinfo {author} {\bibfnamefont {J.}~\bibnamefont
  {P.~Hirth}}\ and\ \bibinfo {author} {\bibfnamefont {J.}~\bibnamefont
  {Lothe}},\ }\href@noop {} {\emph {\bibinfo {title} {{Theory of
  Dislocations}}}}\ (\bibinfo  {publisher} {Krieger Publishing Company},\
  \bibinfo {year} {1982})\BibitemShut {NoStop}%
\end{thebibliography}%

\end{document}